\documentclass[review]{cas-sc} 
\usepackage[authoryear]{natbib}

\usepackage{graphicx}
\usepackage{xspace}

\bibpunct{(}{)}{;}{a}{}{,}
\usepackage{hyperref}
\hypersetup{colorlinks=true, linktocpage=true, pdfstartpage=3,
            pdfstartview=FitV, breaklinks=true, pdfpagemode=UseNone,
            pageanchor=true, pdfpagemode=UseOutlines, plainpages=false,
            bookmarksnumbered, bookmarksopen=true, bookmarksopenlevel=1,
            hypertexnames=true, pdfhighlight=/O, urlcolor=RoyalBlue,
            linkcolor=RoyalBlue, 
            }

\newcommand{\source}[1]{\textsuperscript{\textcolor{blue}{[citation needed]}}\xspace}

\usepackage[switch]{lineno}

\begin{document}
\let\WriteBookmarks\relax
\def\floatpagepagefraction{1}
\def\textpagefraction{.001}
\shorttitle{Asteroid Resurfacing Mechanisms}
\shortauthors{DeMeo, F. E. et~al.}

\title [mode = title]{Isolating the mechanisms for asteroid surface refreshing}

\author[mit]{Francesca E. DeMeo}[orcid=0000-0002-8397-4219] \cormark[1]
\author[mit,eso]{Micha\"el Marsset}[orcid=0000-0001-8617-2425]
\author[wei]{David Polishook}[orcid=0000-0002-6977-3146]
\author[mit,low]{Brian J. Burt}[orcid=0000-0002-6423-0716]
\author[mit]{Richard P. Binzel}[orcid=0000-0002-9995-7341]
\author[jaxa]{Sunao Hasegawa}[orcid=0000-0001-6366-2608]
\author[fin,ltu]{Mikael Granvik}[orcid=0000-0002-5624-1888]
\author[low]{Nicholas A. Moskovitz}[orcid=0000-0001-6765-6336]
\author[mit]{Alissa Earle}[orcid=0000-0002-2780-7037]
\author[ifa]{Schelte J. Bus}[orcid=0000-0003-4191-6536]
\author[nau]{Cristina A. Thomas}[orcid=0000-0003-3091-5757]
\author[apl]{Andrew S. Rivkin}[orcid=0000-0002-9939-9976]
\author[mit]{Stephen M. Slivan}[orcid=0000-0003-3291-8708]

\address[mit]{Department of Earth, Atmospheric, and Planetary Sciences, Massachusetts Institute of Technology, 77 Massachusetts Avenue, Cambridge, MA 02139 USA}
\address[eso]{European Southern Observatory (ESO), Alonso de Cordova 3107, 1900 Casilla Vitacura, Santiago, Chile}
\address[wei]{Faculty of Physics, Weizmann Institute of Science, Rehovot 0076100, Israel}
\address[low]{Lowell Observatory, 1400 West Mars Hill Road, Flagstaff, AZ 86001, USA}
\address[jaxa]{Institute of Space and Astronautical Science, Japan Aerospace Exploration Agency, 3-1-1 Yoshinodai, Chuo-ku, Sagamihara, Kanagawa 252-5210, Japan}
\address[fin]{Department of Physics, University of Helsinki, PO Box 64, FI-00014 Helsinki, Finland}
\address[ltu]{Asteroid Engineering Lab, Space Systems, Lule\aa{} University of Technology, Box 848, S-98128 Kiruna, Sweden}
\address[ifa]{Institute for Astronomy, University of Hawaii, 2860 Woodlawn Drive, Honolulu, HI  96822-1839, USA}
\address[nau]{Northern Arizona University, Department of Astronomy and Planetary Science PO Box 6010, Flagstaff, AZ 86011 USA}
\address[apl]{The Johns Hopkins University Applied Physics Laboratory, Laurel, MD, USA}

\cortext[cor1]{Corresponding author, fdemeo@mit.edu}

\begin{abstract}
Evidence is seen for young, fresh surfaces among Near-Earth and Main-Belt asteroids even though space-weathering timescales are shorter than the age of the surfaces. A number of mechanisms have been proposed to refresh asteroid surfaces on short timescales, such as planetary encounters, YORP spinup, thermal degradation, and collisions. Additionally, other factors such as grain size effects have been proposed to explain the existence of these ``fresh-looking'' spectra. To investigate the role each of these mechanisms may play, we collected a sample of visible and near-infrared spectra of 477 near-Earth and Mars Crosser asteroids with similar sizes and compositions - all with absolute magnitude H$>$16 and within the S-complex and having olivine to pyroxene (ol/(ol+opx)) ratios $>$0.65. We taxonomically classify these objects in the Q (fresh) and S (weathered) classes. We find four trends in the Q/S ratio: 1) previous work demonstrated the Q/S ratio increases at smaller sizes down to H$\lesssim$16, but we find a sharp increase near H$\sim$19 after which the ratio decreases monotonically.  2) in agreement with many previous studies, the Q/S ratio increases with decreasing perihelion distance, and we find it is non-zero for larger perihelia $>$1.2AU, 3) as a new finding our work reveals the Q/S ratio has a sharp, significant peak near $\sim$5$^{\circ}$ orbital inclination, and 4) we confirm previous findings that the Q/S ratio is higher for objects that have the possibility of encounter with Earth and Venus versus those that don't, however this finding cannot be distinguished from the perihelion trend. No single resurfacing mechanism can explain all of these trends, so multiple mechanisms are required. YORP spin-up scales with size, thermal degradation is dependent on perihelion, planetary encounters trend with inclination, perihelion and MOID, noting that asteroid-asteroid collisions are also dependent on inclination. It is likely that a combination of all four resurfacing mechanisms are needed to account for all observational trends.
\end{abstract}


\begin{highlights}
\item We analyze visible and near-ir spectra of 477 S-complex NEOs and Mars Crossers
\item We calculate the fresh-to-space-weathered (Q/S) ratio to understand physical processes
\item The Q/S ratio depends on size, perihelion, inclination, and MOID. 
\item No single resurfacing mechanism can explain all of these trends 
\item Four mechanisms are likely: YORP, thermal degradation, planetary encounters, and collisions.

\end{highlights}


\maketitle


\section{Introduction} \label{sec:intro}

Asteroid spectra reflect both compositional information about the surface as well as non-compositional information. Spectra are dependent on factors such as the grain size of the particles on the asteroid surface, phase angle of the observation, temperature, and by Earth observing conditions of the night such as the airmass \citep{1973JGR....78.8507J,1981JGR....86.4571H,Roush1984,1992LPSC...22..313H,1981JGR....86.4571H,2012Icar..220...36S,Reddy2015,2016AJ....152...54V,Marsset2020}.

Additionally, asteroids are exposed to the space environment which can physically and chemically alter their surfaces over time. This process is called space weathering, an umbrella term that includes many processes such as bombardment by high energy particles from the sun, cosmic rays, and micrometeorite bombardment \citep{1965NYASA.123..711H, 2001JGR...10610039H}. This process was first studied on the moon, confirmed with Apollo samples \citep[e.g.][]{1970Sci...167..745H}

For decades, studies were undertaken to determine the cause of the spectral slope mismatch between many S-complex asteroids and ordinary chondrite meteorites. Lunar-style space weathering was established as playing an important role, reddening the slopes over time, as seen from asteroid spectral studies, laboratory weathering experiments, and ground-truth of sample return from asteroid Itokawa by the JAXA Hayabusa spacecraft \citep[e.g.,][]{Binzel2001,2006Natur.443...56H,Brunetto2006,Nakamura2011}. Different space weathering effects are seen on other asteroid types such as on Vesta \citep{2012Natur.491...79P}. 

Near Earth Objects (NEOs), Mars Crossers (MCs), and small Main Belt Asteroids (MBAs) often display fresh “unweathered” surfaces with low spectral slopes and deep 1-$\mu$m bands due to olivine and pyroxene, spectrally classified as Q-type asteroids. Because space weathering acts on short timescales \citep[$<$1 My, ][]{Vernazza2009}, one could ask: why would we find any fresh surfaces at all? A number of mechanisms have been proposed to rejuvenate asteroid surfaces.

\textit{Planetary Encounters:}
Numerous studies of the orbital trends of Q-types, particularly perihelion distance and the Minimum Orbit Intersection Distance \citep[MOID, ][]{2000A&A...360..411B} to the terrestrial planets suggested close planetary encounters (within $\sim$5-16 planetary radii), particularly with Earth and Venus, are the dominant mechanism for surface refreshing \citep{Nesvorny2005, Marchi2006, Binzel2010, Nesvorny2010}. While Q-types were discovered that encounter only Mars, but not Earth and Venus \citep{DeMeo2014Q}, it was shown that S-type and Q-type asteroids have similar probabilities of an encounter with Mars, while Q-types statistically have more encounters with Venus and Earth than S-types \citep{2016Icar..268..340C}. Additionally, while an increase of Q-types was found at low MOIDs for Venus and Earth, no excess was seen for Mars, further emphasizing the negligible role Mars plays in surface refreshing \citep{2019AJ....158..196D}. 

\textit{YORP spin up:}
Planetary encounters cannot be the sole explanation for the existence of fresh Q-type surfaces.  First, Q-types exist that do not have orbital histories that allow planetary encounters, including among Main Belt Objects. Second, the effects of planetary encounters are not size dependent yet the relative abundance of Q-types increases at smaller sizes \citep[][]{Binzel2004,MotheDiniz2008,Rivkin2011,Thomas2011,Thomas2012,2015Icar..254..202L,2016Icar..268..340C,Graves2018}. \citet{Graves2018} found the change in slope among S-complex asteroids as a function of size was consistent among NEOs, MCs, and MBAs suggesting a common mechanism. 

The Yarkovsky–O’Keefe–Radzievskii– Paddack effect (YORP)  describes the process where thermal emission from an irregularly-shaped body can produce a torque that changes the spin rate of that object over time, causing the body to spin up or spin down \citep{2000Icar..148....2R}. If the spin rate becomes fast enough, material from the surface moves to the equator or is shed from the body, exposing fresh material and even causing the body to fracture \citep{2008Natur.454..188W, 2015MNRAS.454.2249H}. 
\citet{Graves2018} created a resurfacing model from YORP spin-up that could qualitatively explain this decrease in average slope (more Q-type-like) with decreasing size. It could also explain the existence of fresh surfaces at higher perihelia - because the rate of YORP spin-up and space weathering are both dominated by solar radiation there should be no orbital dependence on refresh rates, as discussed in \citet{Graves2018}.

Further support for YORP spinup as a resurfacing mechanism comes from observation of young asteroid pairs. A number of observations have been made of young asteroid pairs in the main belt that were formed by rotational fission, meaning the progenitor body's spin rate was increased by YORP to a point where the body split apart into fragments on similar orbits \citep{2012Icar..221...63M,2014Icar..233....9P,2019Icar..333..165M}. A Q-type asteroid was found among those pairs \citep{2014Icar..233....9P}.

\textit{Collisions and small impacts:} 
\citet{2005Icar..179..325R} modeled the effects of impact-induced seismic activity on asteroid surfaces. They found that small impacts could globally modify the surface including reducing craters. These impacts would also plausibly reset the space weathering age of the surface by exposing fresh material \citep{2013Icar..225..781S}.
 
 \citet{Rivkin2011} and \citet{DeMeo2014Q} estimated with rough order-of-magnitude calculations that the timescale of small impacts able to cause seismic activity that globally resets the regolith on small bodies, both among NEOs and MBAs, is compatible with estimates of the timescale of space weathering. 
 
Both YORP and collisions should be more effective at surface refreshing at smaller sizes. However, the effect of YORP is inclination independent, while collisions would be much more effective for NEOs that have lower inclinations and spend a significant portion of their lifetime in the main belt. \citet{2011MNRAS.416L..26D} found a correlation between the degree of space weathering of an NEO and its probability and intensity of collision.

\textit{Thermal Degradation:} Thermal degradation, also called thermal fatigue or thermally induced surface degradation, is a process by which the diurnal temperature variation on a body breaks up rocks on the surface, and is expected to be the dominant process for regolith creation on small asteroids \citep{2014Natur.508..233D} and even complete disintegration of bodies at low perihelia \citep{2016Natur.530..303G}. Larger boulders experience higher stress than smaller ones, and the effect seems minimal for rock sizes below 30 cm \citep{2017Icar..294..247M}.

\citet{Graves2019} modeled the resurfacing of asteroids due to close planetary encounters and to thermally induced surface degradation, assuming a power law relationship between the resurfacing timescale and the solar distance for degradation.  They find that resurfacing from close encounters cannot reproduce the observed spectral slope trend with perihelion distribution. They also find a much better fit for thermal degradation and suggest that thermal processes are the best explanation for resurfacing asteroids with low perihelia (q $\le$ 0.9 AU).

\textit{Cases against resurfacing mechanisms:}
\citet{MotheDiniz2010} found that over visible wavelengths there are ordinary chondrite spectra that match S-complex asteroids that have higher slopes. They suggest that many S-complex asteroids are actually unweathered.  

\citet{2001JGR...10610039H} suggested that surfaces with less fine regolith (larger grains or boulders) would present a less weathered spectrum. Additionally, \citet{Hasegawa2019} show through laboratory irradiation experiments that for larger-grained ($>$100$\mu$m) samples the spectral slopes are neutral, suggesting that Q-type asteroids could be weathered bodies dominated by large grains on the surface. They find that the space weathering timescales are shorter than the timescales for planetary encounters so as much as half of the Q-type population cannot be fully explained by encounters, and suggest some Q-types have large-grained weathered surfaces.  Contrary to these findings, Itokawa is covered with regolith on parts of its surface and with boulders on other parts, and both regolith-rich and boulder-rich areas present very similar reflectance spectra \citep{2006Sci...312.1334A}. Regions with little fine regolith were found to be highly weathered \citep{2007M&PS...42.1791I}. The relationship between composition, grain size, and space weathering is complex and requires further analysis.

The main scientific objective of this work is to resolve the nature of these most fundamental physical processes that shape the surfaces of the asteroid population. We seek to reduce the spectral effects caused by composition to focus on the effects of space weathering. We use a sample of 477 S-complex NEOs and Mars Crossers with absolute magnitudes (H) greater than 16 and that have high olivine content (ol/(ol+opx) $\ge$ 0.65) consistent with L- and LL-chondrites to maintain compositional homogeneity. We explore the fresh-to-weathered (Q/S) ratios of populations of bodies based on their orbital and physical characteristics as well as their orbital history and potential for encounter with planets.

\section{Data}
 We include visible and near-infrared spectra of 477 S-complex NEOs and Mars Crossers (defined here as perihelion $<$1.67 AU) with H$\ge$16, signal-to-noise ratio at 1$\mu$m $\ge$ 20 and olivine to pyroxene ratio $\ge$0.65 (described in Sec~\ref{sec:shkuratov}) most of which are publicly available from the SMASS and MITHNEOS programs \citep{Bus2002a, Binzel2004, DeMeo2009taxo, Binzel2019, 2022AJ....163..165M} as well as other published literature \citep{1991plas.rept..135W,1998Icar..133...69H,2001Icar..151..139B,Binzel2001,2001PhDT.......121W,Bus2002a,Binzel2004,2004Icar..169..373L,2004MPS...39..351B,Rivkin2004,2005MNRAS.359.1575L,2006AA...451..331M,2007Icar..192..469R,2009ATel.2323....1H,2009Icar..200..480B,2010AA...517A..23D,2010ATel.2571....1H,2010ATel.2822....1H,2010ATel.2859....1H,2011ATel.3678....1H,2011PDSS..145.....H,2012ATel.4251....1H,2013Icar..225..131S,2014AA...569A..59I,Kuroda2014,2014Icar..228..217T,2016AJ....151...11P,2016Icar..268..340C,Binzel2019,2022AJ....163..165M}. The full list of asteroids with their observation dates and visible and near-infrared data references are provided in the Supplementary Data. Most visible-wavelength observations were made between August 1993 and March 1999 using the 2.4-m (f/7.5) Hiltner or 1.3-m (f/7.6) McGraw–Hill reflecting telescopes at the MDM Observatory. Most near-infrared data were obtained using SpeX on the NASA IRTF \citep{Rayner2003}. Data collected through October 2021 were included in the analysis. New observations taken between October 2020 and October 2021 (SpeX runs sp273 through sp286) are newly published in this work. Observation and reduction details for SpeX data can be found in \citet{2022AJ....163..165M}. A number of asteroids had near-infrared spectral measurements from multiple nights, in which case we keep the spectrum with the highest signal-to-noise ratio. The sample includes only objects with H$>$16 which corresponds roughly to bodies $<$2 km across. We restrict the sample because most Q-types are found below that size. Spectral plots of the 477 asteroid spectra are provided in the Supplementary Data. 

\section{Methods}

\subsection{Taxonomic Classification of Spectra} \label{sec:classification}
Most data were given taxonomic assignments in previous work \citep{Bus2002a, Binzel2004, DeMeo2009taxo, Binzel2019, 2022AJ....163..165M}, and we use the existing classifications for the spectra used in this work. New data were classified using the Bus-DeMeo classification system \citep{DeMeo2009taxo} (a classification tool is publicly available, \url{smass.mit.edu/busdemeoclass.html}) and visual inspection of spectral features as described in detail in \citet{2022AJ....163..165M}. We include all S-complex objects in this work except for Sa-types which have much higher olivine contents and are not compositionally equivalent to ordinary chondrites \citep{1984Sci...223..281C,2007M&PS...42..155S}.

We then more carefully distinguish between Q-types and the rest of the S-complex. For asteroids for which we have data over the full 0.5-2.45$\mu$m range, we use the unambiguous classification result from the Bus-DeMeo system. For asteroids with near-infrared-only data in the 0.8-2.5$\mu$m range for which the classification system does not provide a unique result, we use the method from \citet{DeMeo2014Q} to determine the appropriate designation within the S-complex or Q-type.

For the analysis in this work we focus on the ratio of fresh-to-weathered asteroids, which is represented taxonomically as the ratio of Q-types to S-types (Q/S ratio). As is done in \citet{DeMeo2014Q}, a weighting of 0.5 is given to objects assigned as ``Q:'' (degenerate between Sq and Q). Any object classified as S, Sq, or Sr is grouped as ``S''. Any object classified as Q is kept as Q.

\subsection{Mineralogic Modeling: Shkuratov Method} \label{sec:shkuratov}
 We performed a mineralogical analysis of S-complex spectra using the \citet{Shkuratov1999} radiative transfer model. The asteroid spectra were reddened using the empirical function from \citet{Brunetto2006}, using their space weathering factor ``Cs'' as a free parameter in the fitting process. Our spectra were modeled following the procedure described by \citet{2008Natur.454..858V} and \citet{Binzel2019}.  The output product of our Shkuratov modeling is an estimate for the ol/(ol+px) ratio that provides the best fit to the spectrum out to 1.9$\mu$m. Model results are provided in Supplementary Material.
 
 Because Q-types have spectra that are olivine-rich, similar to L- and LL-chondrites, we remove any S-complex spectra with low olivine to pyroxene ratios (ol/(ol+opx) $<$0.65) from the analysis. The resulting dataset includes 477 asteroids.

\subsection{Minimum Orbit Intersection Distance Calculation} \label{sec:moid}
We calculate the orbit and Minimum Orbit Intersection Distance (MOID) to Venus, Earth, and Mars over the past 500,000 years. This is done in the same manner as \citet{Binzel2010} and \citet{DeMeo2014Q} using the \texttt{swift\_rmvs3} code from \citet{Levison1994} with a 3.65-day time step and output values computed at 50-year intervals, accounting for the eight planets Mercury to Neptune. Our integrations for each asteroid included six additional clones, test particles with the same initial position, offset with velocities differing by $\pm 6 \times 10^{-6}$ astronomical unit (AU) per year in each Cartesian component. These clones are important to evaluate the range of potential orbital evolutions. From these orbital histories we find the MOID for Venus, Earth, and Mars.

We then define a number of ``MOID classes'' based on the planets for which an asteroid has a close approach. We define a ``close approach'' as a MOID smaller than 10 planetary radii of the given planet. The MOID classes include: V+E+M (encounters with Venus, Earth, and Mars), V+E (Venus and Earth), Earth, E+M (Earth and Mars), Mars, and None (No encounters with planets).  Asteroids in the ``None'' class are unlikely to have encountered a planet over the 500,000 year integration.  Among a sample of 225 NEOs that were unlikely to have encountered a planet (``None'' MOID class) over the past 500,000 years, we ran a longer integration test of 2,000,000 years and found 218 (97\%) remained in that class.

\begin{figure}
	\centering
		\includegraphics[width=0.5\textwidth]{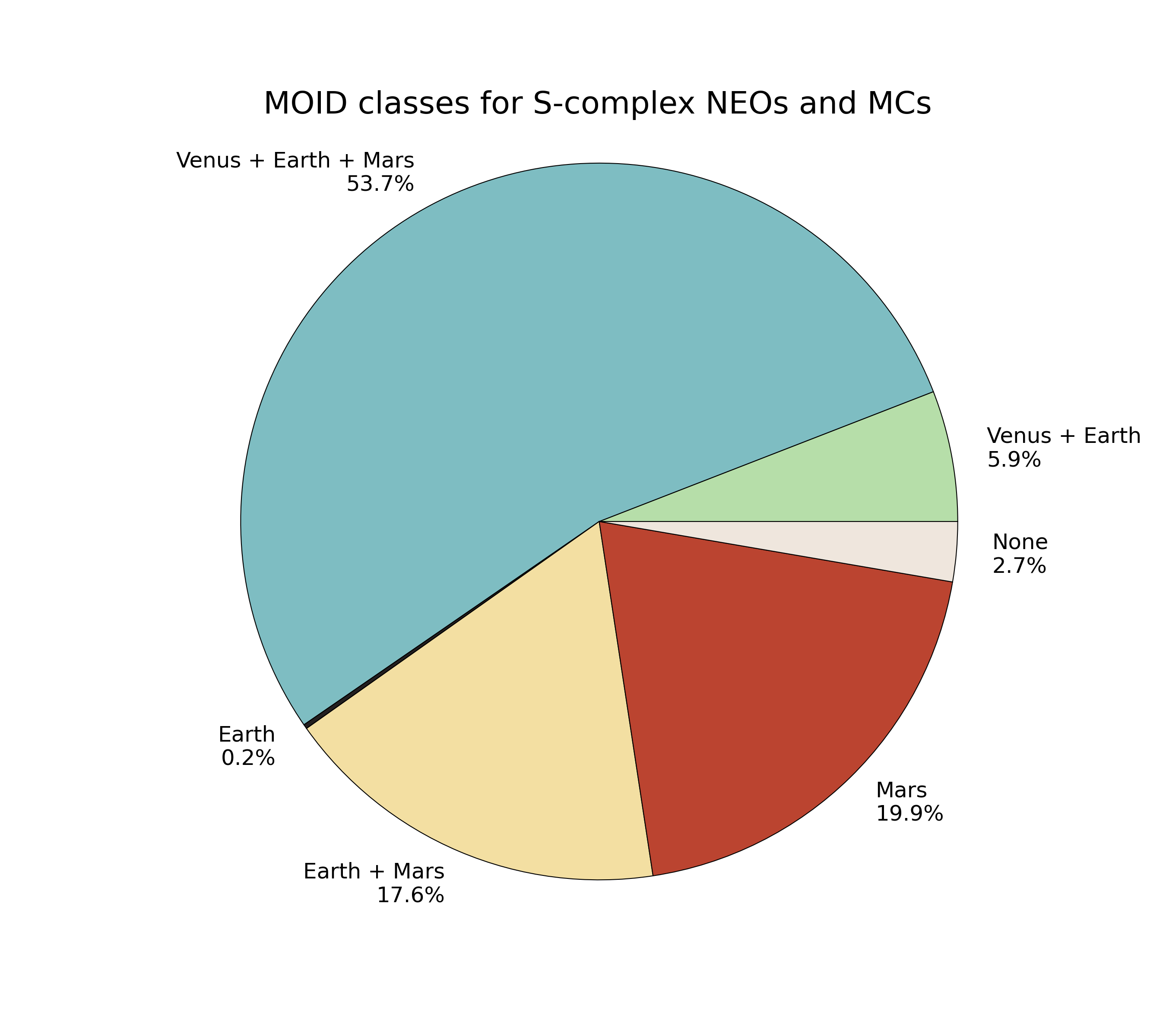}
	\caption{The number of NEOs and MCs that fall into each MOID class. For example, 54\% of our sample (256 objects) encounter the orbits of Venus, Earth and Mars (V+E+M class), whereas 3\% of our sample (13 objects) do not encounter any of the planets (None class).}
	\label{FIG:MOIDclassSummary}
\end{figure}

Half of the sample crossed all planets (V+E+M), the Earth + Mars and Mars-only populations each accounted for $\sim$20\%, and the Venus + Earth and None populations each made up 6\% and 3\% as shown in Fig.~\ref{FIG:MOIDclassSummary}. There were very few asteroids in our sample that only crossed Earth, and none that only crossed Venus.

\section{Results} \label{Sec:Results}

\begin{figure}
	\centering
		\includegraphics[width=0.47\textwidth]{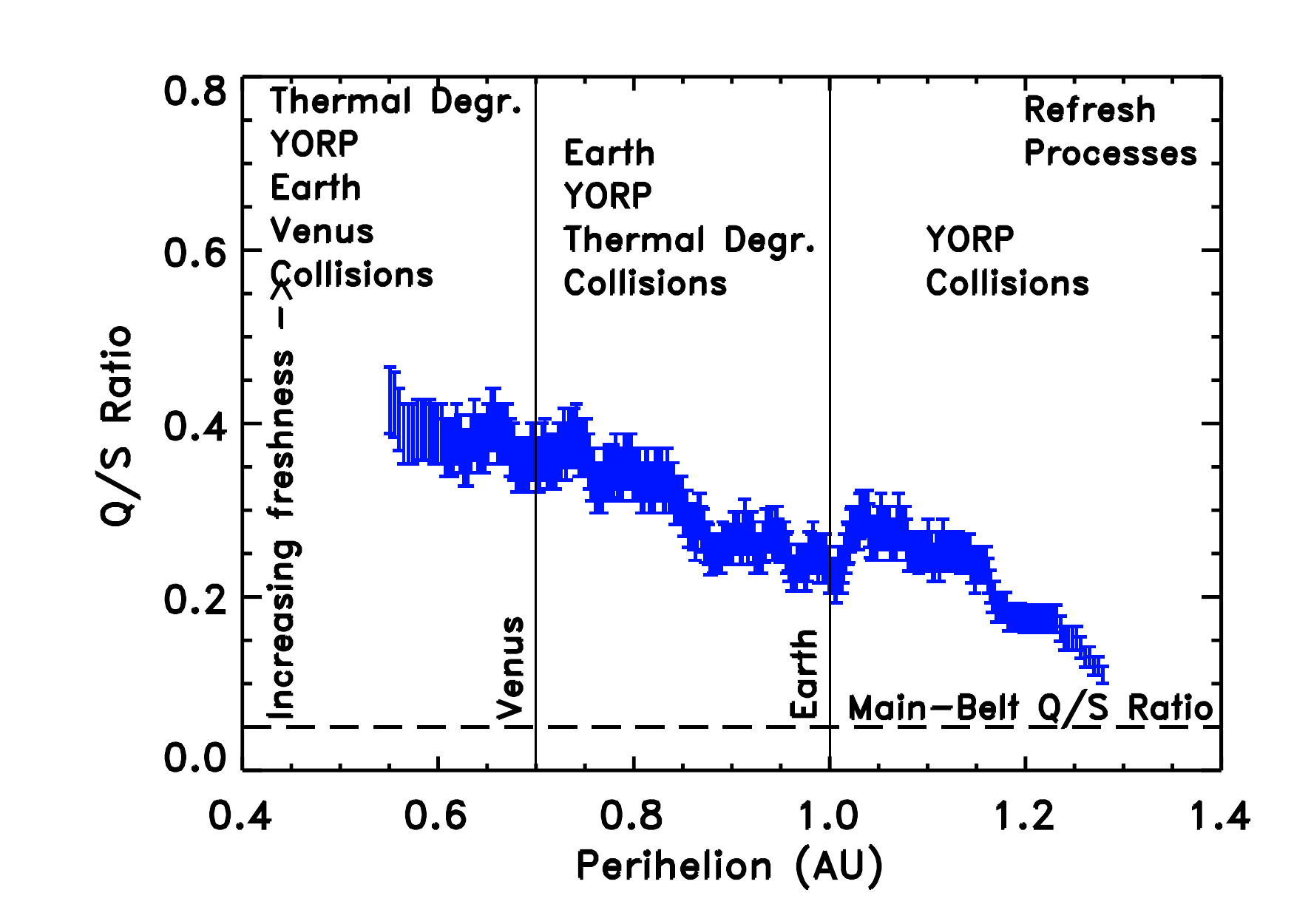}
		\includegraphics[width=0.47\textwidth]{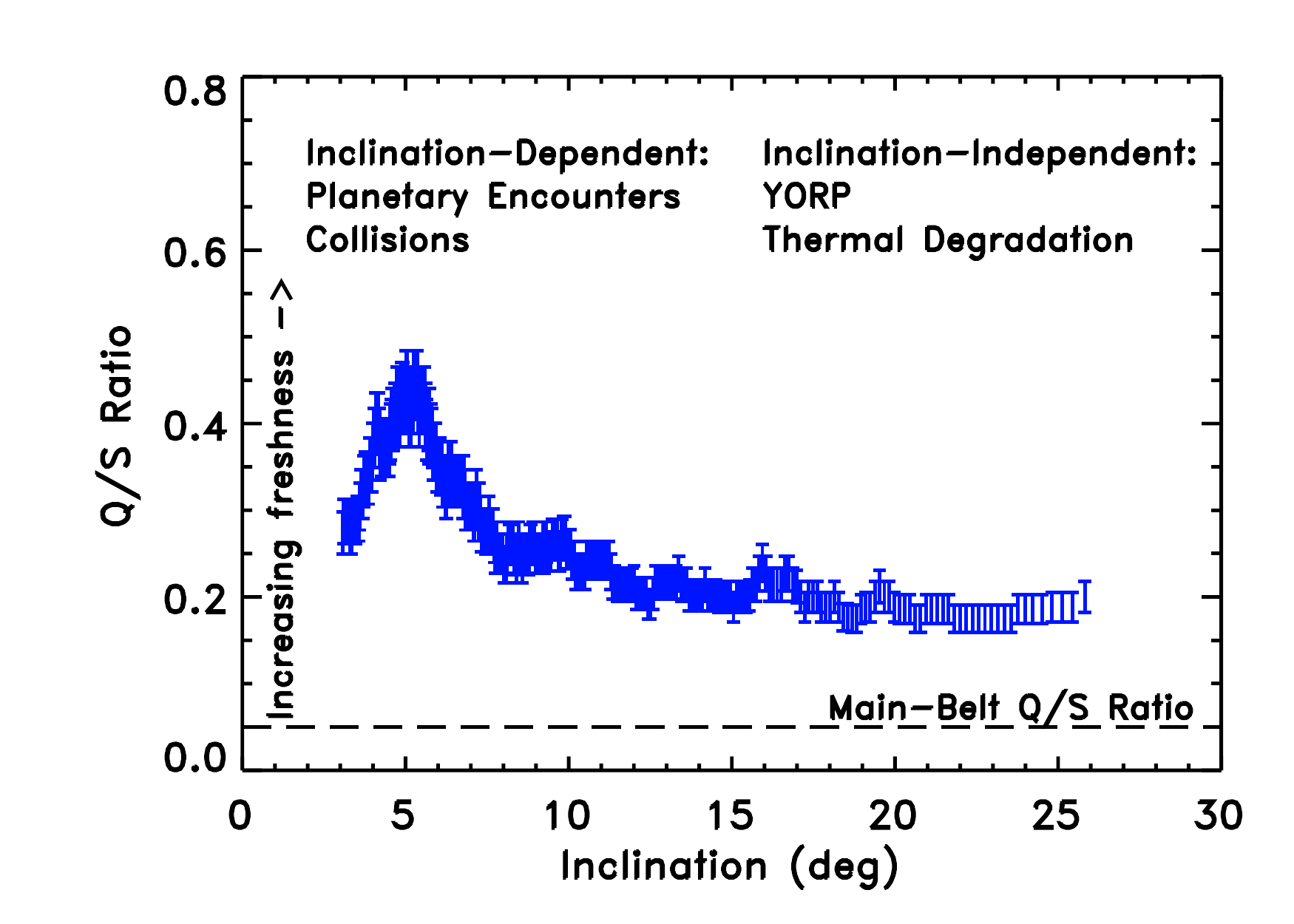}	
		\includegraphics[width=0.47\textwidth]{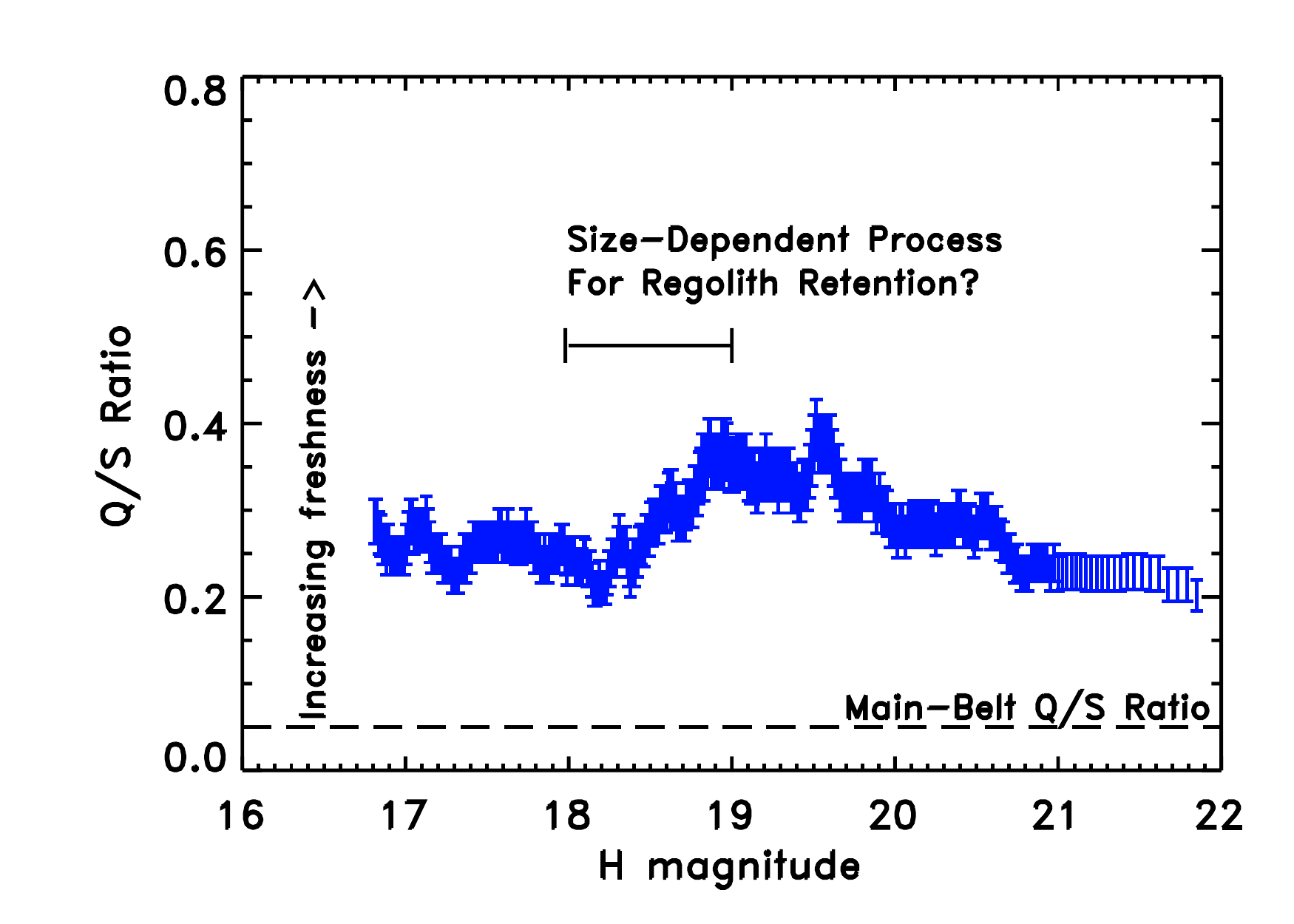}
		\includegraphics[width=0.47\textwidth]{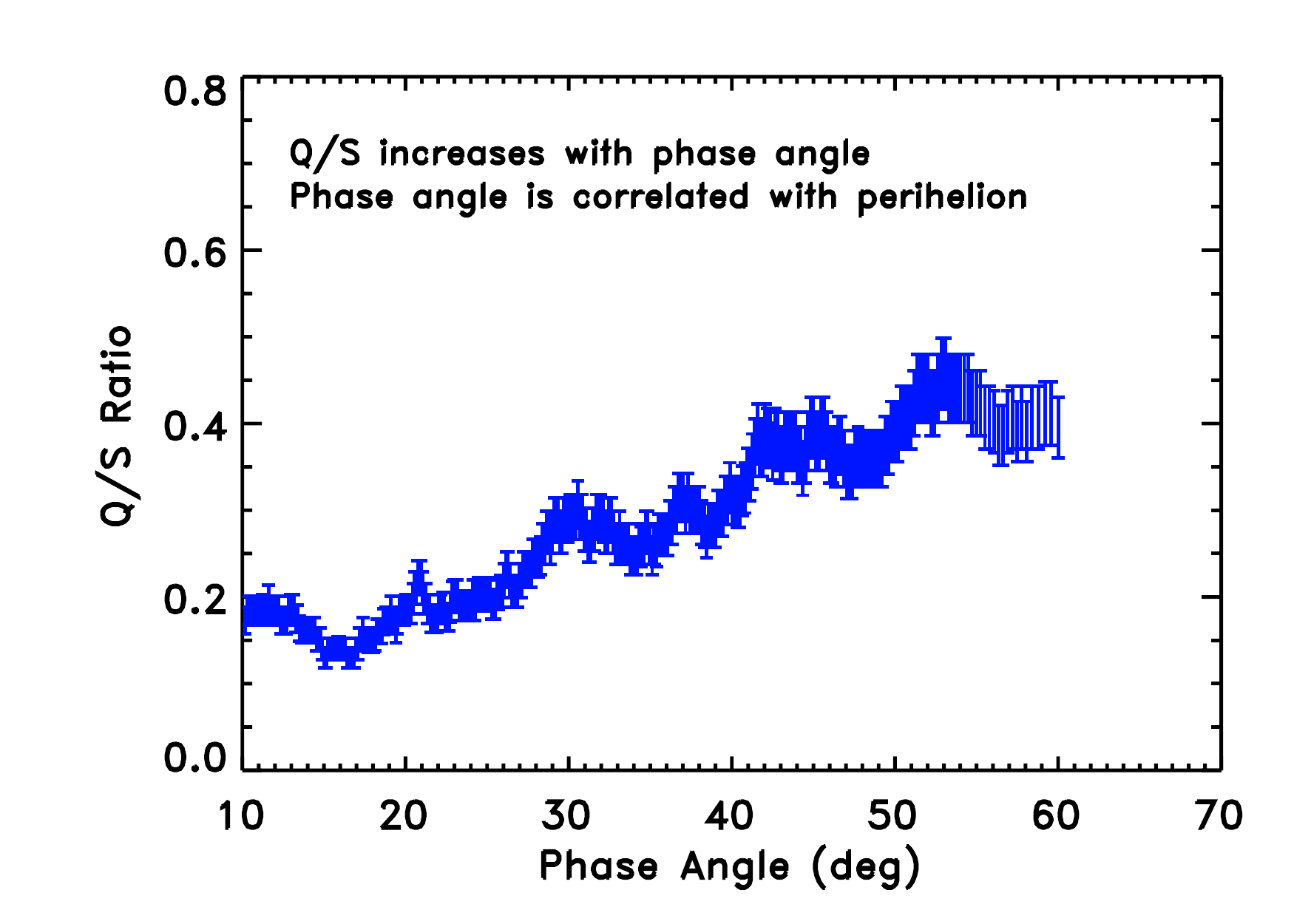}

	\caption{The running Q/S ratio for perihelion, H magnitude, inclination and phase angle for the 477 asteroids with ol/(ol+opx) $\ge$ 0.65 with a box size of 125.  Poisson errors are plotted as the Q/S ratio (y value) divided by the square root of n, where n is number of objects in the bin. The estimated main-belt Q/S ratio is from \citet{2015Icar..254..202L} and represents a similar size range as this dataset. The observational phase angle and perihelion are correlated, with phase angle increasing as perihelion decreases.
	}
	\label{FIG:qhi}
\end{figure}

Using the dataset of 477 S-complex asteroids with ol/(ol+opx) ratios $>$0.65 we explore orbital and physical trends with the Q/S ratio (defined in Sec.~\ref{sec:classification}) to interpret possible refreshing mechanisms. In Fig.~\ref{FIG:qhi} we plot the running Q/S ratio as a function of perihelion, H magnitude, and inclination. Below, we describe the Q/S trend for each of them.

\textit{Q/S Trend by size:}  As shown in Fig.~\ref{FIG:qhi} the Q/S trend with H magnitude is relatively flat over the H magnitude range of 16--18. There is an increase in the Q/S ratio near H magnitude of 19, and for H$\gtrsim$19.5 the Q/S ratio decreases linearly.  The Q/S ratio near the minima and maximum on this plot is 0.29$\pm$0.03 (H=17.0--18.0, 86 objects), 0.45$\pm$0.05 (H=18.7--19.7, 80 objects), and 0.17$\pm$0.02 (H$\ge$21.0, 83 objects). There is a $>$1$\sigma$ difference between the H=17.0--18.0 region and both other regions. There is a $>$3$\sigma$ difference between the maximum at H$\sim$19 and H$>$21.0 (this 3$\sigma$ difference is not visually apparent in Fig.~\ref{FIG:qhi} with the large box size of 125).

\citet{Binzel2004} and \citet{Thomas2012} show decreasing average S-complex slope with size, inferring a higher fraction of Q-types at smaller sizes. Using data from the Sloan Digital Sky Survey \citep[SDSS,][]{2001AJ....122.2749I}, \citet{Thomas2012} analyze Main Belt Objects in the Koronis family up to H$\sim$15.3, so there is no overlap with our data that start at H $>$16. \citet{Graves2018} found a similar trend in the Flora family. The observations from \citet{Binzel2004} span from 10 km to 0.4 meters, but the slope decrease happens mostly for diameters greater than 1 km (H=16--17) and levels off at smaller sizes. Our result is consistent with these works, showing that overall, there is an increase of fresher asteroids at smaller sizes down to H magnitude roughly 16, but we find a sharper increase in the Q/S ratio near H$\sim$19 and a decreasing ratio at smaller sizes.

\textit{Q/S Trend by inclination:} We find the Q/S ratio peaks between 3$^{\circ}$ and 6$^{\circ}$, seen as the strong peak in the running Q/S ratio, centered at i $\sim$5$^{\circ}$ in Fig.~\ref{FIG:qhi}. The Q/S ratio drops off steeply at both higher and lower inclinations. The Q/S ratio is 0.17$\pm$0.02 between i=0--3$^{\circ}$ (59 objects), 0.45$\pm$0.04 between i=3--6$^{\circ}$ (99 objects), and 0.25$\pm$0.02 between i=6--10$^{\circ}$ (122 objects). There is a 4$\sigma$ difference between the i=0--3$^{\circ}$ and i=3--6$^{\circ}$ groups and a 3$\sigma$ difference between the i=3--6$^{\circ}$ and i=6--10$^{\circ}$ groups.

\textit{Q/S Trend by perihelion:}
We see a clear trend of an increasing Q/S ratio with decreasing perihelion, as seen by many previous authors \citep[e.g.,][]{Marchi2006,2019AJ....158..196D,2019Icar..324...41B}. We find no Q-types with perihelia > 1.3 AU although our sample is relatively small at that distance (34 objects).

\textit{Q/S Trend by phase angle:} 
We also see a clear trend of increasing Q/S with increasing phase angle, as was seen by \citet{DeMeo2014Q}. This trend initially seems counter-intuitive because increasing phase angle should increase the spectral slope making an object less likely to be classified as Q-type. We find no correlation with phase angle and asteroid size. We find that phase angle and perihelion are correlated, as phase angle increases, perihelion decreases. Because the number of Q-types increases with decreasing perihelion, it is natural that as a by-product the average phase angle at which those bodies are observed increases.

\begin{figure}
	\centering
		\includegraphics[width=0.8\textwidth]{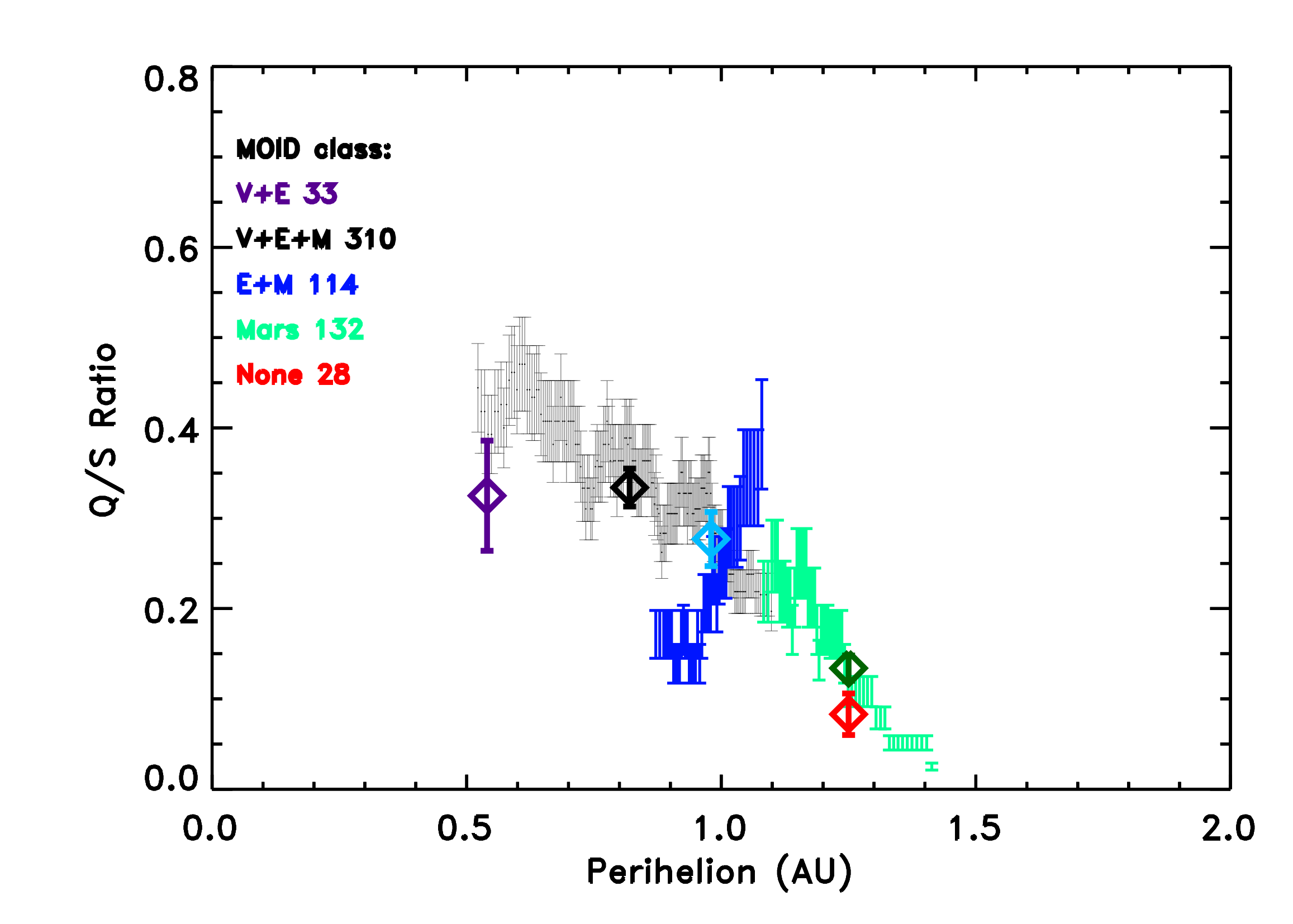}
	\caption{The running Q/S ratio with perihelion displayed by MOID class. Each class is plotted in a unique color as shown on the plot legend, and the number of objects in each class is also listed. Black includes objects that cross Venus, Earth, and Mars. Blue includes objects that cross Earth and Mars, and in green are objects that only cross Mars. The running mean box size is 81 for the V+E+M population and 41 for the others. We plot as a diamond the overall average Q/S ratio and perihelion for each MOID class including two additional groups that did not have enough data for a running ratio: the Venus + Earth population in purple and the None (crossing no planet) population in red.}
	\label{FIG:perimoid}
\end{figure}

\textit{Q/S Trend by MOID class:} In Fig.~\ref{FIG:perimoid} we plot the Q/S ratio by perihelion for each MOID class (defined by the  planets the object could encounter, see Sec.~\ref{sec:moid}).  We see a trend where encounters with planets (including more and larger planets) has a higher Q/S ratio, but the overall perihelion trend appears to dominate these results. We note a few interesting trends:

\begin{itemize}
\item First, the Earth+Mars population has the opposite trend from the other classes and from the overall trend -- there is a peak in the Q/S ratio near 1.1 AU and a minimum near 0.9 AU. There are 114 objects in this group. However, The average Q/S ratio of this population follows the overall perihelion trend.
\item Second, the average Q/S ratio for the Mars class is 0.136$\pm$0.014 and for the None class is 0.083$\pm$0.023. The ratios of the two classes differ by more than 1$\sigma$ (using Poisson uncertainties) but less than 2$\sigma$. This result emphasizes that the role of Mars in surface refreshing is either small or zero \citep{2016Icar..268..340C,2019AJ....158..196D}. 
\item Third, because the ``None'' population does have some Q-types, it represents a definitive baseline for Q-types created by processes other than planetary encounters. 
\item Fourth, the V+E (Venus- and Earth-crossing) average Q/S ratio is lower than V+E+M (Venus-, Earth- and Mars-crossing) ratio at the lowest perihelion distances, however, the two overall class averages are very similar.
\end{itemize}

\textit{Trends by source region:}
To look for trends as a function of source region, we use probabilities for an asteroid to enter the NEO region through a certain escape region---defined as a combination of mean-motion and/or secular resonances or as a group separated from other asteroids in orbital-element space---from the main asteroid belt as predicted by the model of \citet{2018Icar..312..181G}. We note that, for the resonance complexes, the probabilities do not necessarily correspond to the true source regions, but to the last location of a given asteroid just before it enters the NEO region. \citet{2022AJ....163..165M} calculated and published probabilities for most of the asteroids considered here and we have calculated the probabilities for the remaining ones using the same approach. Of the asteroids in our sample with high olivine content (ol/(ol+opx) $\ge$ 0.65), 359 had $>$50\% probability of entering the NEO region through the $\nu_6$ resonance complex, representing 75\% of our sample. 302 objects or 63\% had $>$70\% $\nu_6$ probability.  Of all the other escape regions considered by the model (the Hungaria, Phocaea, and JFC groups, and the 3:1, 5:2, and 2:1 resonance complexes) there were a negligible number of asteroids with $>$70\% probability of coming from that escape region (less than 10 objects for each region). Hence we assign each of these NEOs that have a high olivine content the probability of originating from the $\nu_6$ resonance complex. 

\begin{figure}
	\centering
		\includegraphics[width=0.47\textwidth]{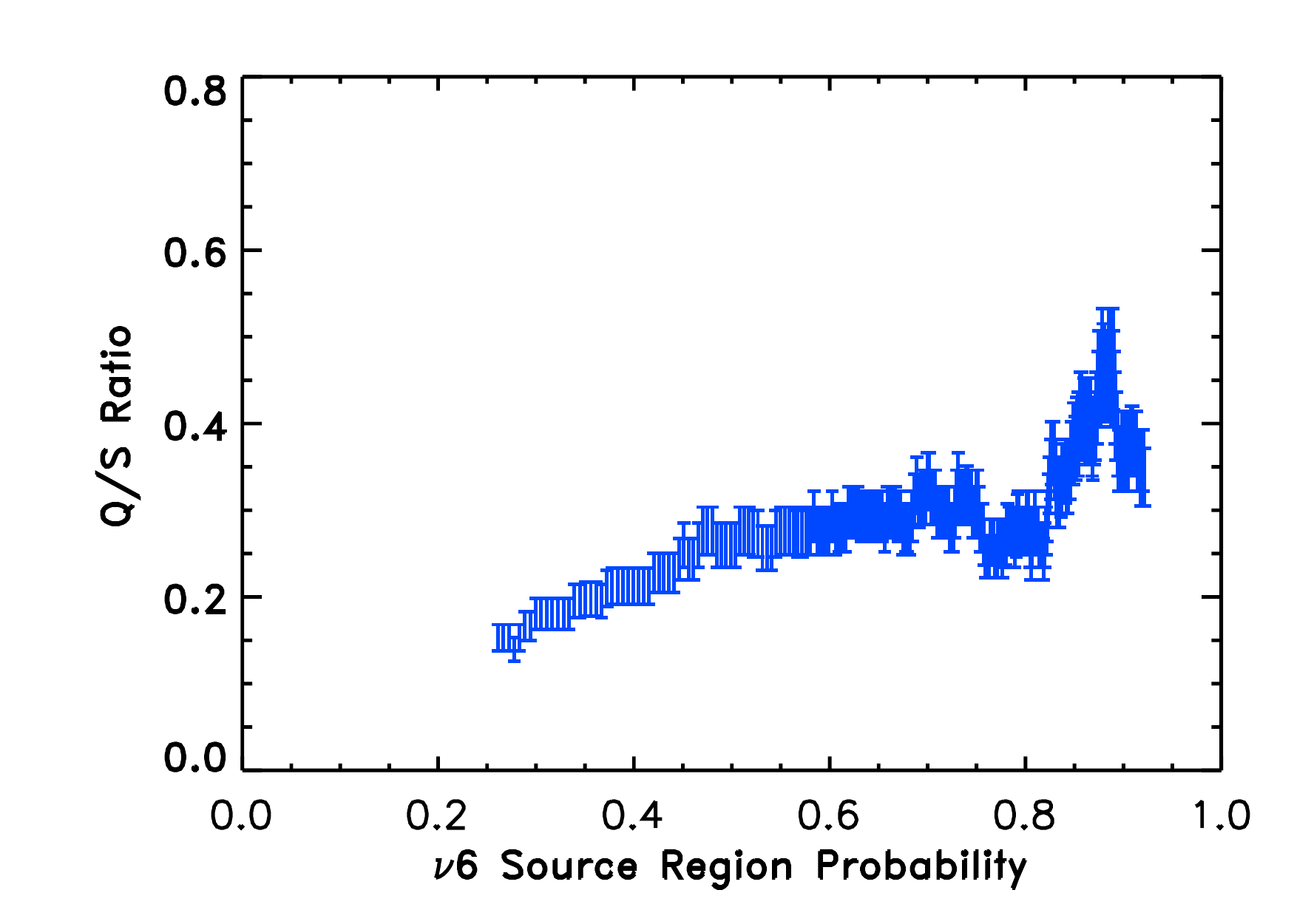}
		\includegraphics[width=0.47\textwidth]{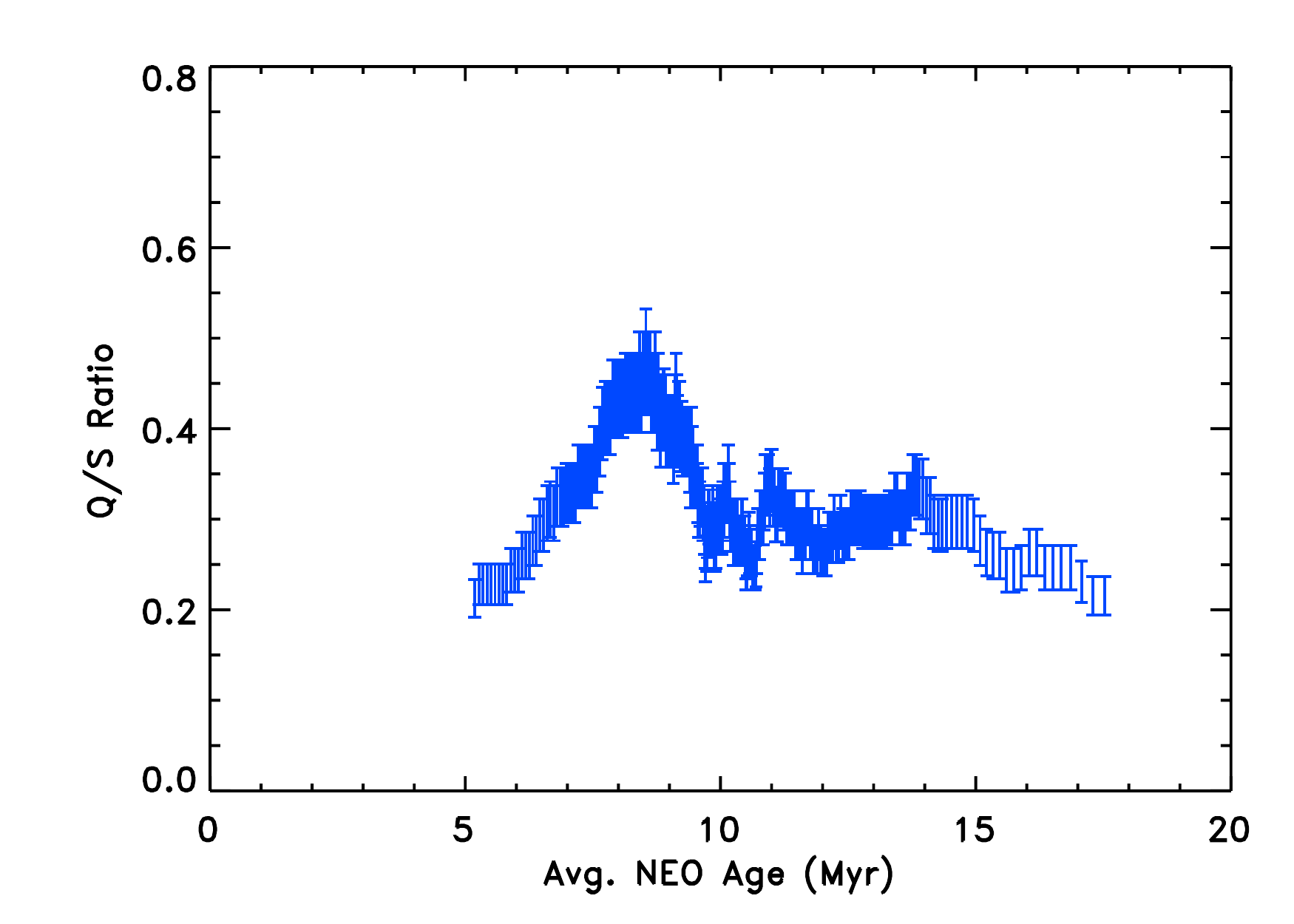}
	\caption{
	Left: The running Q/S ratio over the probability of being sourced from the $\nu_6$ secular resonance. There is a spike in the Q/S ratio (more fresh bodies) for $\nu_6$ source probabilities $>$80\%. Right: The Q/S ratio as a function of average NEO age calculated as the sum of a body's probability from each source region time the mean NEO lifetime from that region from \citet{2018Icar..312..181G}. The mean life time from the $\nu_6$ is 9.4 and 7.2My for perihelia less than and greater than 1.3 AU, respectively. The peak in Q/S ratio near 8 My corresponds with the higher probability of originating from the $\nu_6$.
	}

	\label{FIG:srage}
\end{figure}

In Fig.~\ref{FIG:srage} we plot the Q/S ratio as a function of increasing $\nu_6$ probability and find as the $\nu_6$ probability increases, the Q/S ratio increases. However, there is a strong spike in the Q/S ratio for $\nu_6$ probabilities $>$80\%. We also calculate an average age for each NEO by multiplying each source probability by the mean NEO lifetime from that source from \citet{2018Icar..312..181G}. We find a peak in the Q/S ratio near 8 My corresponds with the higher probability of originating from the $\nu_6$ (mean lifetime between 7.2 and 9.4 My).

\section{Discussion} \label{Sec:Discussion}
\subsection{Discussion of Primary Q/S Trends}
There are 4 primary trends with the Q/S ratio. Each refreshing mechanism (described in the Introduction, Sec.~\ref{sec:intro}) is dependent on different factors, such as size, distance from the Sun, and close encounters with planets, and the dominant refreshing mechanisms should be able to explain one or more of these trends. We summarize the observed trends and dependencies in Table~\ref{Table:trendsummary} and discuss these trends in the context of potential refreshing mechanisms.

\begin{table*}[width=1\linewidth,pos=h]
\caption{Q/S Trends with Orbital and Physical Properties}
\label{Table:trendsummary}
\begin{tabular}{p{0.08\textwidth}p{0.3\textwidth}p{0.12\textwidth}p{0.12\textwidth}p{0.3\textwidth}}
\toprule
Property	&	Observed Q/S Trend	&	Dependent Mechanism	&	Independent Mechanism	&	References	\\
\midrule					\hline			
Size	&	Dependent: Q/S increases for decreasing size for H$<$16, peaks near H$\sim$19 and decreases at smaller sizes	&	YORP	&	Encounters, Collisions, Thermal Degradation	& \citet{Binzel2004, MotheDiniz2008,Rivkin2011,Thomas2011,Thomas2012,2015Icar..254..202L,2016Icar..268..340C,Graves2018}, This Work.		\\
\hline
Inclination	&	Dependent: Q/S peaks at 3--6$^{\circ}$	&	Encounter, Collisions	&	YORP, Thermal Degradation	& \citet{2011MNRAS.416L..26D}, This Work.		\\
\hline
Perihelion	&	Dependent: Q/S increases for q$\lesssim$0.9 AU. Independent for q$\gtrsim$0.9AU.	&	Thermal Degradation, Encounter, Collisions	&	YORP	&	\citet{Marchi2006,2019AJ....158..196D,2019Icar..324...41B}, This Work.	\\
\hline
MOID	&	Dependent? MOID effects not easily distinguished from perihelion effects &	Encounter	&	YORP, Collisions, Thermal Degradation	&	\citet{Nesvorny2010,Binzel2010,DeMeo2014Q,2016Icar..268..340C,2019AJ....158..196D}, This Work.	\\

\bottomrule
\end{tabular}
\end{table*}

First, the Q/S ratio increases at smaller sizes down to H$\lesssim$16, peaks at H$\sim$19 and then decreases with increasing H. The lowest Q/S ratio is seen at the smallest sizes, H$\ge$21.0. YORP is size-dependent and is more effective at smaller sizes, particularly in the kilometer size range and smaller \citep{Bottke2006}. Collisions are dependent on size to some extent as well. While the rate of collisions is roughly size-independent (in fact it increases with cross-sectional area or size), the larger the body the less effective a micro-impact will be at global surface refreshing. Planetary encounters and thermal degradation should not be size dependent. \citet{Graves2018} modeled the spectral slope versus size (H magnitude) trend for asteroids for planetary encounters and YORP (modeling refreshing happening at the fission spin rate, though it could happen more often). They found YORP could adequately model the spectral slope trend with H-magnitude trend, but planetary encounters could not.

We find a peak in the Q/S ratio near H$\sim$19  (see Fig.~\ref{FIG:qhi}). One possible explanation is that we are seeing the transition between ``large'' NEOs which are all rubble piles and smaller bodies some of which are monoliths that can surpass the spin-barrier limit \citep{2000Icar..148...12P,2009Icar..202..134W}. As the number of regolith-free monoliths increases (at smaller sizes), the Q/S ratio would decrease. Additionally, with decreasing size, decreasing gravity decreases regolith retention. There are two competing forces between monolith creation from fission from YORP spinup, and regolith creation from thermal degradation.

Second, the Q/S ratio increases with decreasing perihelion distance. Thermal degradation should scale as a power law with distance from the Sun, meaning more fresh surfaces at lower perihelia. \citet{Graves2019} modeled thermal degradation as a method of reducing spectral slope (refreshing a surface) and was able to adequately reproduce the observed trend. The YORP effect should be distance independent because both YORP and space weathering are dependent on solar radiation and should thus scale at the same rate with distance  \citep{Graves2018}.  Collisions are most effective for bodies on orbits that spend more time in the Main Belt, and so would not cause the observed perihelion trend. Planetary encounters are indirectly related to perihelion, but require a unique MOID signature with the planets to identify it. At lower perihelia more (perhaps all) processes are at work, and thus the combined forces would also naturally increase the abundance of fresh surfaces. Each process may have a different physical refreshing effect. Thermal fatigue takes over to break down boulders, creating new, fresh regolith at a very fast rate. YORP could be massively shedding outer layers, exposing fresh surface. Planetary encounters may be overturning existing, weathered grains, exposing fresh material.

Third, our analysis reveals that the Q/S ratio is significantly higher at inclinations between 3 and 6$^{\circ}$ than at lower and higher inclinations.  We are able to dismiss the processes of YORP and thermal degradation, as these have no dependence on inclination.  We have three possible interpretations of this trend: 

\begin{itemize}
\item Planetary encounters are very strongly coupled to inclination, where the dynamical likelihood of planetary encounters scales with 1/sin(i) \citep{1967JGR....72.2429W}.  
For asteroids evolved out of the main belt into planet crossing orbits, planetary encounters dominate over asteroid-asteroid collisions \citep{Nesvorny2010}.  Planetary encounters, however, are most efficient for bodies on orbits with inclinations near zero. If the signal we see at low inclinations (i$\sim$5$^{\circ}$)  is due to planetary encounters then, the drop for i$\lesssim$5$^{\circ}$ indicates depletion of the population either by collision or frequent encounters that fling the bodies into a new orbit (higher i).  At i=0$^{\circ}$ planet crossing collision is effectively inevitable. The sample cannot peak near 0$^{\circ}$ as the survival time near 0 is too short. 

\item The inclination signal is an indicator of NEO average lifetime. The median inclination of the inner belt is around 5.2$^{\circ}$ (as calculated from MPCORB.dat from the Minor Planet Center in May 2022). The Flora family is also centered near 3--6$^{\circ}$ \citep{2015aste.book..297N}.  Flora is a major source of NEOs through the $\nu_6$ resonance, we also see a spike in the Q/S ratio for objects with very high probability ($>$85\%) of coming through the $\nu_6$ (Fig.~\ref{FIG:srage}). NEOs from the $\nu_6$ have a much longer mean lifetime (7-9 My) than NEOs originating from the J3:1 (2 My) or J5:2 (0.05 My) \citep[average NEO lifetimes from ][]{2018Icar..312..181G}. The longer lifetime indicates more time for resurfacing processes (any or all of them) to take place.

\item Collisions: \citet{2011MNRAS.416L..26D} found a correlation between NEO slopes and the estimated intensity of collisional processes. While those authors do not note a direct correlation between slope and inclination, they show collisions are more frequent for low inclinations and for aphelia of about 2.75 AU.  In a survey of 150 small MBAs, \citet{2015Icar..254..202L} find the Q/S ratio to be less than 5\% down to an observational limit of ~1km.  The relative paucity of Q-types among small MBAs indicates that while asteroid-asteroid encounters create some resurfacing, this process operates at a very low background level.  The Flora family, specifically, is a source of S-type NEOs \citep{2002Icar..157..155N}, but is not a major source of small meteorite-sized bodies \citep{2008Natur.454..858V} -- only 80\% of meteorite falls are ordinary chondrite, but only 10\% of falls are LL-type (Flora-like compositions). Both the low fraction of Q-types among the Flora family and the paucity of small meteorite-sized bodies (meter size or smaller to cause refreshing but not total disruption) from Flora argues against there being a significant population of small bodies in that family causing refreshing collisions among NEOs.

\end{itemize}

Fourth, the Q/S ratio is higher for objects that have the possibility of encounter with Earth and Venus versus Mars or no planet. This trend is more challenging to interpret uniquely because it strongly overlaps with the perihelion trend. Objects that interact with Venus and Earth naturally have lower perihelia on average than those that do not. Previous work that found Q-types had higher probabilities of interacting with planets than S-types \citep{Binzel2010, 2016Icar..268..340C} were biased because they included higher perihelia S-types \citep{Graves2019}. Additionally, \citet{Graves2018} and \citet{Graves2019} could not satisfactorily model the observed H magnitude and perihelion trends by planetary encounters. The sharp decrease in fresh surfaces for objects in the Earth + Mars MOID class in Fig.~\ref{FIG:perimoid} is curious, but may be a signal of the dynamical influx from the $\nu_6$ resonance, where first-time encounters are particularly efficient at exposing fresh material.  Bodies in this region are in a very active phase of Earth encounters, where fresh material is frequently exposed, then saut\'eed, until the upper regolith layer is saturated by space weathering.  Alternatively or in addition, repeated encounters could lead to a loss of regolith during an interval where an object is strongly coupled to the planet's orbit.


There is observational evidence that YORP (correlated to H magnitude) and thermal degradation (correlated to perihelion) and collisions (correlated to inclination) each play a role in surface refreshing. The existence of fresh surfaces in the main belt further supports the role of YORP and collisions. Planetary encounters likely plays a role, however, the existing orbital trends can be explained by other refreshing mechanisms, and we are unable to uniquely distinguish encounters as a mechanism in this work. We see support for the role of planetary encounters in the peak in the Q/S ratio at low inclinations near 5$^{\circ}$ (Fig.~\ref{FIG:qhi}).

\subsection{Discussion of Caveats and Other Explanations}
\textit{Compositional Homogeneity Caveat:} Space weathering effects vary based on the composition of the body. In this work, we have attempted to use a relatively compositionally homogenous sample that includes S-complex bodies with high olivine content (ol/(ol+opx) $\ge$0.65). Additionally, a large portion of this sample is likely sourced from the $\nu_6$ resonance which is also dominated by the large Flora family. However, it is still possible, that there are biases in our trends based on different populations of NEOs being sourced from different regions of the of the Main Belt.

\textit{Grain Size Alternative Explanation:} How significant is the effect of grain size on an asteroid surface? \citet{Hasegawa2019} performed irradiation experiments of ordinary chondrite meteorites of varying grain sizes and found that the spectral slopes remained low for large-grained irradiated samples. They suggest that a significant fraction of Q-types could simply be weathered surfaces with large grains.

Some smaller asteroids are monoliths and may not have a significant regolith layer \citep{2000Icar..148...12P}. Perhaps some smaller bodies have surfaces dominated by 30-cm boulders, the size limit at which thermal degradation is not longer effective.  Spectra of larger grains and slabs typically have weaker band depths compared to smaller grains \citep[e.g.,][]{Cloutis2011,KramerRuggiu2021,2022Icar..38014971D}, yet Q-type asteroids have both low spectral slopes and deep 1$\mu$m absorption bands. An investigation of other band parameters other than slope such as width and depth would should help constrain the role grain size plays in our interpretation of Q-type asteroids and their relative fraction among the S-complex.

\section{Conclusion}

We find 4 trends with the Q/S ratio representing fresh versus weathered surfaces with size, perihelion, inclination, and MOID. The complexity of these trends highlights that no single resurfacing mechanism can explain all of these trends, so multiple mechanisms are required. 

These observational trends are:
\begin{itemize}
    \item Size: The Q/S ratio increases at smaller sizes down to H$\lesssim$16, then peaks near H$\sim$19 after which the ratio decreases, revealing that there are fewer fresh surfaces at the smallest sizes, particularly smaller than H$>$21. 
    \item Perihelion: The Q/S ratio increases with decreasing perihelion. Our results and previous work have shown that the Q/S ratio is small, but non-zero for perihelia $>$1.2 AU, including among Mars Crossers and the Main Asteroid Belt.
    \item Inclination: The Q/S ratio exhibits a strong peak for inclinations between 3--6$^{\circ}$.
    \item MOID: The Q/S ratio is higher for objects that have the possibility of encounter with Earth and Venus versus Mars or no planet (low MOID).  The major trends seen for MOID, however, can be easily confounded with the trend for perihelion.
\end{itemize}

The possible explanations for the observed trends (see Table~\ref{Table:trendsummary}):
\begin{itemize}
    \item YORP may dominate the observed trend with size \citep{Graves2018}, but not with perihelion or inclination.
    \item Thermal degradation likely dominates the observed trend with perihelion \citep{Graves2019}.
    \item Planetary encounters are likely the dominant mechanism to explain the observed trend with inclination. Encounters could also explain the perihelion trend and Q/S ratio for objects with low MOIDs, however, most low perihelion asteroids (of any taxonomic class including Q or S) also have low MOIDs making it difficult to separate the effect of MOID from perihelion.
    \item Collisions could be responsible for producing Q-types at low inclinations and also at large distances such as in the Main Belt. They could also be size dependent because smaller bodies would be more likely to be globally resurfaced by a micro-impact compared to a larger body, although this has not yet been modeled at the population level.
    \item Grain size could play a role, with larger grains causing lower spectral slopes regardless of weathering level. Additional investigation of band parameters such as width, depth, and area would provide further constraints.
\end{itemize}

\section{Acknowledgements}
  We thank David Nesvorn\'y for helpful discussions that are always insightful. We also thank Vishnu Reddy and an anonymous referee for helpful reviews. Observations reported here were obtained at the NASA Infrared Telescope  Facility,  which is operated by the University of Hawaii under under contract 80HQTR19D0030  with the National Aeronautics and Space Administration.   The MIT component of this work is supported by NASA grant 80NSSC18K1004 and 80NSSC18K0849.
  SH was supported by the Hypervelocity Impact Facility (former name: The Space Plasma Laboratory), ISAS, JAXA.  Any opinions, findings, and conclusions or recommendations
  expressed in this article are those of the authors and do not necessarily reflect the views of the National Aeronautics and Space Administration.




\bibliographystyle{cas-model2-names}

\bibliography{resurface}


\end{document}